\documentclass[twocolumn,twoside,slac_two]{revtex4}
\usepackage{graphicx}
\usepackage{fancyhdr}
\newcommand{\pizero}{\mbox{$\pi^{0}$}}
\newcommand{\pizeroll}{\mbox{$K_{L} \to \pi^{0} l l$}}  
\newcommand{\klong}{\mbox{$K_{L}$}}
\newcommand{\kshort}{\mbox{$K_{S}$}}
\newcommand{\ktopipi}{\mbox{$K \to \pi \pi$}} 
\newcommand{\pitoeeee}{\mbox{$\pizero \to e^{+}e^{-}e^{+}e^{-}$}}
\newcommand{\pitoeeg}{\mbox{$\pizero \to e^{+}e^{-}\gamma$}}
\newcommand{\ktopimue}{\mbox{$\klong \to \pizero \mu^{\pm} e^{\mp}$}}
\newcommand{\ktopipimue}{\mbox{$\klong \to \pizero \pizero \mu^{\pm} e^{\mp}$}}
\newcommand{\ktomumuee}{\mbox{$\klong \to \mu^{\pm}\mu^{\pm} e^{\mp}e^{\mp}$}}
\newcommand{\pitomue}{\mbox{$\pizero \to \mu^{\pm} e^{\mp}$}}
\newcommand{\ktomue}{\mbox{$\klong \to \mu^{\pm} e^{\mp}$}}
\newcommand{\mutoeg}{\mbox{$\mu^{\pm} \to e^{\pm} \gamma $}}
\newcommand{\mutoe}{\mbox{$\mu^{\pm} + N \to e^{\pm} + N $}}
\newcommand{\rk}{\mbox{$R_{K}$}}
\newcommand{\rkdef}{\mbox{$\rk = \frac{\Gamma\left(K^{\pm} \to e^{\pm} \nu_{e} \right)}{\Gamma\left(K^{\pm} \to \mu^{\pm} \nu_{\mu}\right)}$}}

\newcommand{\ptsq}{\mbox{$p_{T}^{2}$}}
\newcommand{\ktopipipid}{\mbox{$\klong \to \pizero \pizero \pizero_{D}$}}
\newcommand{\ktopipipi}{\mbox{$\klong \to \pizero \pizero \pizero$}}
\newcommand{\kltopiee}{\mbox{$\klong \to \pizero e^{+} e^{-}$}}

\newcommand{\kstopiee}{\mbox{$\kshort \to \pizero e^{+} e^{-}$}}
\newcommand{\ktopigg}{\mbox{$\klong \to \pizero \gamma\gamma $}}
\newcommand{\ktopieeg}{\mbox{$\klong \to \pizero e^{+} e^{-}\gamma $}}
\newcommand{\av}{\mbox{$A_{V}$}}
\newcommand{\ktopizpiz}{\mbox{$\klong \to \pizero \pizero $}}
\newcommand{\ktopizpizd}{\mbox{$\klong \to \pizero \pizero_{D} $}}
\newcommand{\mkaon}{\mbox{$M_{K}$}}
\newcommand{\kethree}{\mbox{$K_{e}3$}}
\newcommand{\kefour}{\mbox{$K_{e}4$}}

\begin{document}
\title{New Rare Decay Results from KTeV}

\author{Michael Akashi-Ronquest for the KTeV collaboration}
\affiliation{Department of Physics and Astronomy \\
University of North Carolina, Chapel Hill}

\begin{abstract}
Data from the KTeV experiment have provided many opportunities for studies of fundamental symmetries in nature, not just CP. 
We present new limits on the lepton flavor violating decays \ktopimue, \ktopipimue\ and \pitomue. We also present measurements of \ktopigg\ and \ktopieeg\
which are relevant to the determination of the various contributions to \pizeroll. Finally, we present an analysis of the decay \pitoeeee.
\end{abstract}

\maketitle

\section{Introduction}
Symmetries have yielded a significant number of insights into the interactions of fundamental particles. While CP violation has been intensely studied for signs of new
physics, only recently has lepton flavor violation (LFV) exposed cracks in the Standard Model. It is not presently clear what kind of mechanism is responsible for the violation of
lepton flavor symmetry, and the situation is not helped by the fact that the violation of this symmetry has thus far appeared in only the neutrino sector, where experiments 
are extremely challenging. New limits on other manifestations of LFV may be very useful in constraining models, while the kaon sector provides a landscape where
backgrounds to such searches are quite manageable. However, heavily suppressed CP violating decays, such as \pizeroll, remain sensitive probes of physics beyond the Standard
Model owing to the ease with which the SM predictions can be produced. Finally, CPT remains one of the few conserved symmetries, and as such is an excellent place in which to
look for new physics.  

\section{The KTeV Experiment} 
KTeV is the culmination of a series of neutral kaon experiments at Fermilab. The project consisted of two different experiments sharing a
common detector. E799-II focused on rare \klong\ decays. E832 was designed to establish the existence of Direct CP
violation in  \ktopipi. The KTeV detector, described more fully in \cite{EprimePRD2003} consisted of a long vacuum decay
volume followed by a magnetic spectrometer, a very high performance CsI crystal calorimeter and finally a muon veto. KTeV, when
operating in E799 configuration, also included a transition radiation detection (TRD) system, designed to improve the detector's already
excellent particle ID capability. In this mode, two nearly parallel \klong\ beams were produced. One the other
hand, the E832 configuration inserted an active regenerator, consisting of a series of plastic scintillator blocks instrumented with
PMTs, into the beam. This regenerator alternated between the two kaon beams, and had the effect of transforming the relevant beam into a coherent \klong-\kshort\
beam.  

\begin{figure}[htb]     
	\includegraphics[scale=0.30]{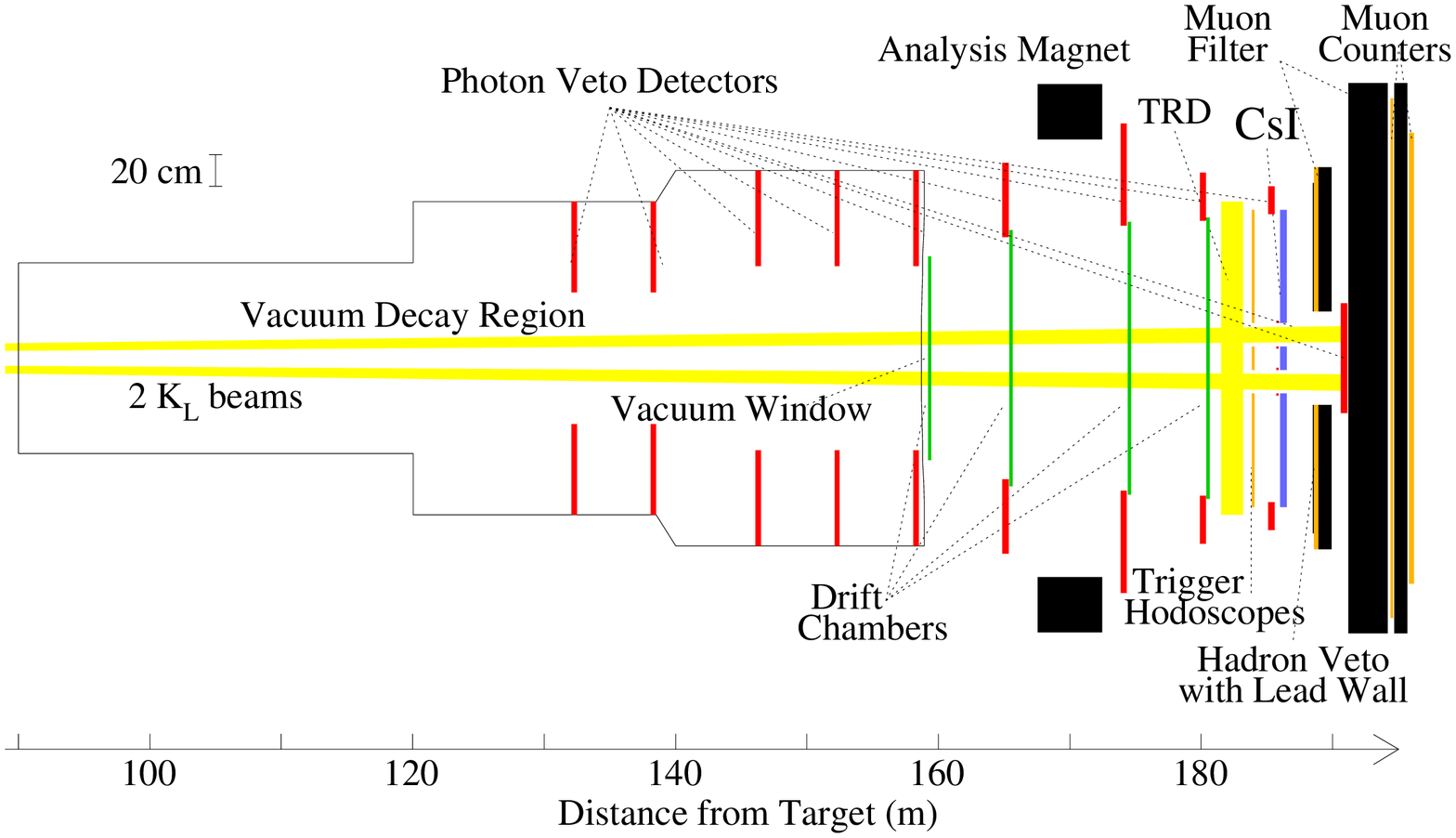}
	\caption[A diagram of the KTeV detector in the E799 configuration]{A cartoon of the KTeV detector in the E799 configuration. The relative positions of the major detector components are shown, including the drift chambers and
	CsI calorimeter. Notice the twin kaon beams entering the detector. 
           }
       \label{pic:e799_diagram}
 \end{figure} 

\section{Search for Lepton Flavor Violation}
With the observation of neutrino oscillation \cite{Ahmad2002} it is now known that lepton flavor is not a perfectly conserved symmetry of nature. With the
root mechanism of neutrino oscillation still unknown, it is important to search for other manifestations of this symmetry violation. 
Neutral kaon decays can offer an alternative probe of LFV. The experimental signature is simple: the
presence of the lepton pair $\mu^{\pm} e^{\mp}$ in any kaon decay. 
It should be emphasized that while these modes could occur due to neutrino mixing, said rates would be undetectable in current experiments \cite{Langacker:1988vz}.
Thus observation of any new LFV process would provide a unique signature of new physics beyond the standard model, distinct from neutrino mixing.

\subsection{Search for \ktopimue}
One decay mode which would exhibit LFV is \ktopimue. The E799 data set is filtered \cite{LFV2008} to look for events in which there is a charged track which matches hits
in the muon detector, implying that a muon is indeed present. The other charged track must then have a CsI cluster whose energy is equal to the momentum as measured using
magnetic deflection, implying that the shower in the CsI was electromagnetic in origin and thus due to an electron.  Additionally, the transistion radiation detector information for the
electron-like track must also be consistent with an electron. Only events with two charged tracks are accepted. Additionally, there must be two more CsI clusters
not associated with any tracks, which come from the two $\gamma$s from the \pizero. With these requirements, the signal is dominated by \kethree\ and/or \kefour\ decays
where the pion either decays (and thus produces an muon) or punches through the CsI calorimeter to the muon detector. Making a tight cut on accidental activity
in the detector eliminates extra CsI clusters which are needed for \kethree\ events to be accepted, while \kefour\ events can be reduced by calculating $|p_\nu|$, the
magnitude of the neutrino momentum assuming that the decay was actually a \kefour\ event, and then retaining events for which  $|p_\nu| < 0$, which is the case when
the \kefour\ hypothesis is incorrect.

\begin{figure}[htb]     
	\includegraphics[scale=0.40]{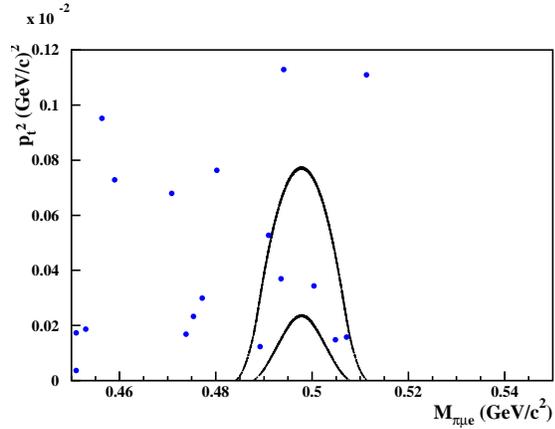}
	\caption{A plot of the invariant $\pizero \mu e$ mass versus transverse momentum squared for the \ktopimue\ sample. The larger (smaller) contour defines
	the region which encloses 99\% (95\%) of \ktopimue\ Monte Carlo events. 
           }
       \label{pic:ktopimue_mass_plot}
 \end{figure}

Using a detailed Monte Carlo simulation of the decay and the detector, we define contours as a function of the $\pizero\mu e$ invariant mass and \ptsq\ ( the
squared component of the summed momentum which is transverse to the neutral kaon beam) which would enclose either 99\% or 95\% of the total \pitomue\ sample.
The signal region is enclosed within the 95\% contour and is blinded while the control region is defined as falling between the 95\% and 99\% contours.
Both contours are shown in Figure  \ref{pic:ktopimue_mass_plot}.

Monte Carlo studies of all possible
sources of backgrounds indicate that the background expected in the control region is $4.21 \pm 0.53$ events while for the control region the expected number is
$0.66 \pm 0.23$ events. Unblinding the signal region, we observe no events, while we observe 5 events in the control region. This leads to a limit on the
branching ratio of
\begin{equation}
Br(\ktopimue) < 7.56 \times 10^{-11}
\end{equation}
at 90\% confidence, which is a factor of 83 lower than the previous limit \cite{Arisaka1998}.

\subsection{Search for \ktopipimue}
Very similar to \ktopimue\ is \ktopipimue. The additional \pizero\ requires the presence of two more CsI photon clusters for reconstruction, the side effect of
which is a strong suppression of the accidental background which was an issue for \ktopimue. Due to the lower accidental background, it is possible to relax cuts
relative to \ktopimue\ in order to raise the sensitivity--- in this case the tight cut on accidental activity in the detector is removed, as is the TRD cut on the
electron track. After applying these cuts, and others as detailed in \cite{LFV2008}, the primary background is \ktopipipid, $\pizero_{D} = \pitoeeg$  events. In order for \ktopipipid\
events to contribute to the background, one electron must fail to deposit a significant amount of energy into the CsI calorimeter, and at the same time, there
must be an accidental hit in the muon detector which matches the electron track's trajectory. In order to prevent this, we place a loose requirement on the TRD information for
the muon-like track, requiring that the track is \emph{not} electron-like. After this cut is added, there are an expected $0.44 \pm 0.23$ background events in the
signal region. In the data, after all cuts, there are no events in the signal region, leading to a branching ratio limit of:
\begin{equation}
Br(\ktopipimue) < 1.7 \times 10^{-10}
\end{equation}
at 90 \% confidence. This is the first limit placed on this decay.

\subsection{Search for \pitomue}
The \ktopipimue\ search can be easily extended by noticing that the decay \ktopipipi, \pitomue\ has the same final state. All that is required to constrain the
rate at which this decay occurs is the additional requirement that the $\mu e$ pair reconstruct near the \pizero\ invariant mass. After enforcing this additional
cut, we are led to the branching ratio limit of:
\begin{equation}
Br(\pitomue) < 3.59 \times 10^{-10}
\end{equation}
at 90 \% confidence. This is a factor of 10 (2) times lower than the previous best limits on $\pizero \to \mu^{-} e^{+}$ ($\pizero \to \mu^{+} e^{-}$)
\cite{Appel2000A,Appel2000B}. Note that this analysis is equally sensitive to both charge modes.

\subsection{Other searches}
While the above neutral kaon decay modes are the complete set of LFV modes that KTeV has recently investigated ( \ktomumuee\ was the first LFV decay studied by KTeV
\cite{AlaviHarati:2002eh} ) it should be noted that this does not represent 
all the possible kaon decay modes which may exhibit lepton flavor violation. One good example is the decay \ktomue. While KTeV's excellent CsI calorimeter resulted in best
current limits on all LFV modes involving neutral pions, a search for this decay was undertaken by the B871 collaboration using a detector optimized for two body decays 
\cite{Ambrose:1998us}. Their resulting branching ratio limit was Br(\ktomue) $< 4.7 \times 10^{-12}$ at 90\% confidence. 

The charged kaon sector also offers excellent probes
for LFV, such as \rkdef . This quantity offers both a very precise standard model prediction \cite{Spadaro2008,Cirigliano:2007ga} and is sensitive to lepton flavor
violating contributions \cite{Testa:2008xz,Masiero:2005wr}. Current measurements of \rk\ have uncertainties in the range of 3\% while preliminary results from KLOE
\cite{Testa:2008xz}, NA48/2 \cite{Spadaro2008} and NA62 (aka NA48/3) \cite{Spadaro2008} 
may reach 1\%, 2\% and 0.5\% respectivley. Both the KLOE and NA62 collaborations plan upgrades to their detectors in an effort to push the precision down to the SM uncertainity
of 0.04\%.   

In addition to searches in the kaon sector, there are other areas in which lepton flavor violation may be observed. One process which is currently under investigation
is  \mutoeg\ ( current limit is Br(\mutoeg) $< 1.2 \times 10^{-11}$ \cite{Ahmed:2001eh} )  which will be probed down to the $10^{-13}$ level by the MEG experiment \cite{Hisamatsu:2007zza}
Another lepton flavor violating process is the related process \mutoe\ in which the transition involves a virtual photon coming from a nucleus instead of an emitted real photon. This process, which would
have been studied by the MECO experiment had the RSVP project proceeded, may instead be investigated by the Mu2e experiment, an evolution of the MECO design which would be sited
at Fermilab \cite{Mu2e2007}. That experiment plans to search for \mutoe\ down to approximately $R_{\mu e} < 10^{-16}$,
where $R_{\mu e}$ is the ratio of the rate of \mutoe\ to the rate of muon capture for the same given nucleus. This will be a significant improvement over the current limit of $4.3 \times
10^{-12}$, obtained using a Ti target \cite{Dohmen:1993mp}.

\subsection{Final Note}
It should be emphasized that all of the lepton flavor violating searches detailed here are background free. Specifically, the estimated background for \ktopimue\
is 0.66, for \ktopipimue\ the estimate is 0.44 and the expected background for \pitomue\ is 0.03 events. This implies that a KTeV style experiment with additional beam
intensity can further lower these limits, especially \pitomue\ due to the multiplicity of neutral pions produced via \ktopipipi.

\section{CP conserving contributions to \pizeroll}
While lepton flavor violation is a direct gateway to new physics, various extensions to the Standard Model result in a multitude of new or enhanced CP violation
effects. The suite of decays that comprise \pizeroll\ are especially attractive targets for further study due to the heavily suppressed standard model predictions 
and the precision with which the SM predictions can be made. 

However, these modes are also a challenge to observe, especially $\klong \to \pizero \nu \bar{\nu}$.
\kltopiee on the other hand offers a higher branching ratio and a fully reconstructible final state, at the cost of a less precise standard model prediction. In
addition, \kltopiee receives contributions from CP conserving, direct CP violating and indirect CP violating terms. The challenge is to try to determine the sizes
of the relative contributions in order to focus on the direct CP violating component, which is most sensitive to new physics effects. Observation of \kstopiee\ constrains
the magnitude of the indirect CP violating term \cite{Batley:2003mu} which then leaves the size of the CP conserving component to be determined. 

\subsection{\ktopigg} 
The vector meson exchange amplitude \av\ can be used to determine the size of the CP conserving component of \pizeroll\ \cite{Gabbiani2002}
In order to produce a sample of \ktopigg\ events, data from the E832 dataset is searched for events containing four CsI clusters, and no charged tracks.
The energy of each of the CsI clusters is required to be above 2 GeV. The distribution of these photon clusters is also required to be consistent with a kaon
decay within the E832 configuration's pure \klong\ beam. The clusters are then paired such that one pair yields an invariant mass within 3 MeV of the mass of the
\pizero, while the other pair is required to \emph{not} reconstruct into a \pizero. This last requirement suppresses the background due to \ktopizpiz.
After these cuts and others are applied \cite{Pi0gg2008} we are left with 1982 events with an expected background level of approximately 30\%.

The decay \ktopizpiz\ was chosen as the normalization mode, as it has the same final state. The resulting branching ratio is:
\begin{equation}
Br(\ktopigg) = \left(1.29 \pm 0.06 \right) \times 10^{-6}
\end{equation}
which is consistent with the result from NA48 \cite{Lai2002} and supersedes the previous KTeV result \cite{Pi0gg1999}. 

\begin{figure}[htb]     
	\includegraphics[scale=0.40]{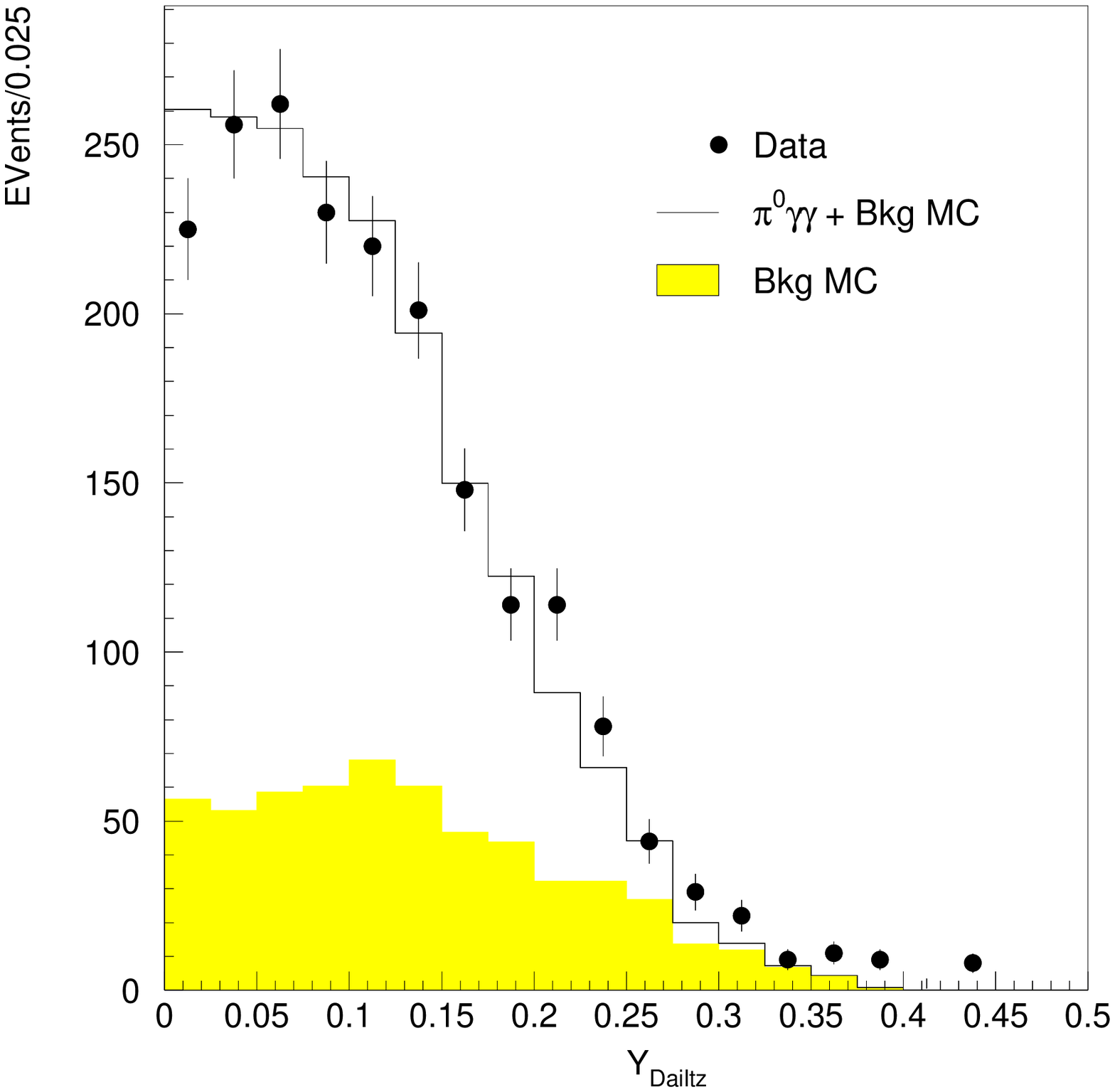}
	\caption{Distribution of the variable $Y_{Dalitz}$ used in the analysis of \ktopigg. 
           }
       \label{pic:ktopigg_ydalitz_plot}
 \end{figure}

\begin{figure}[htb]     
	\includegraphics[scale=0.40]{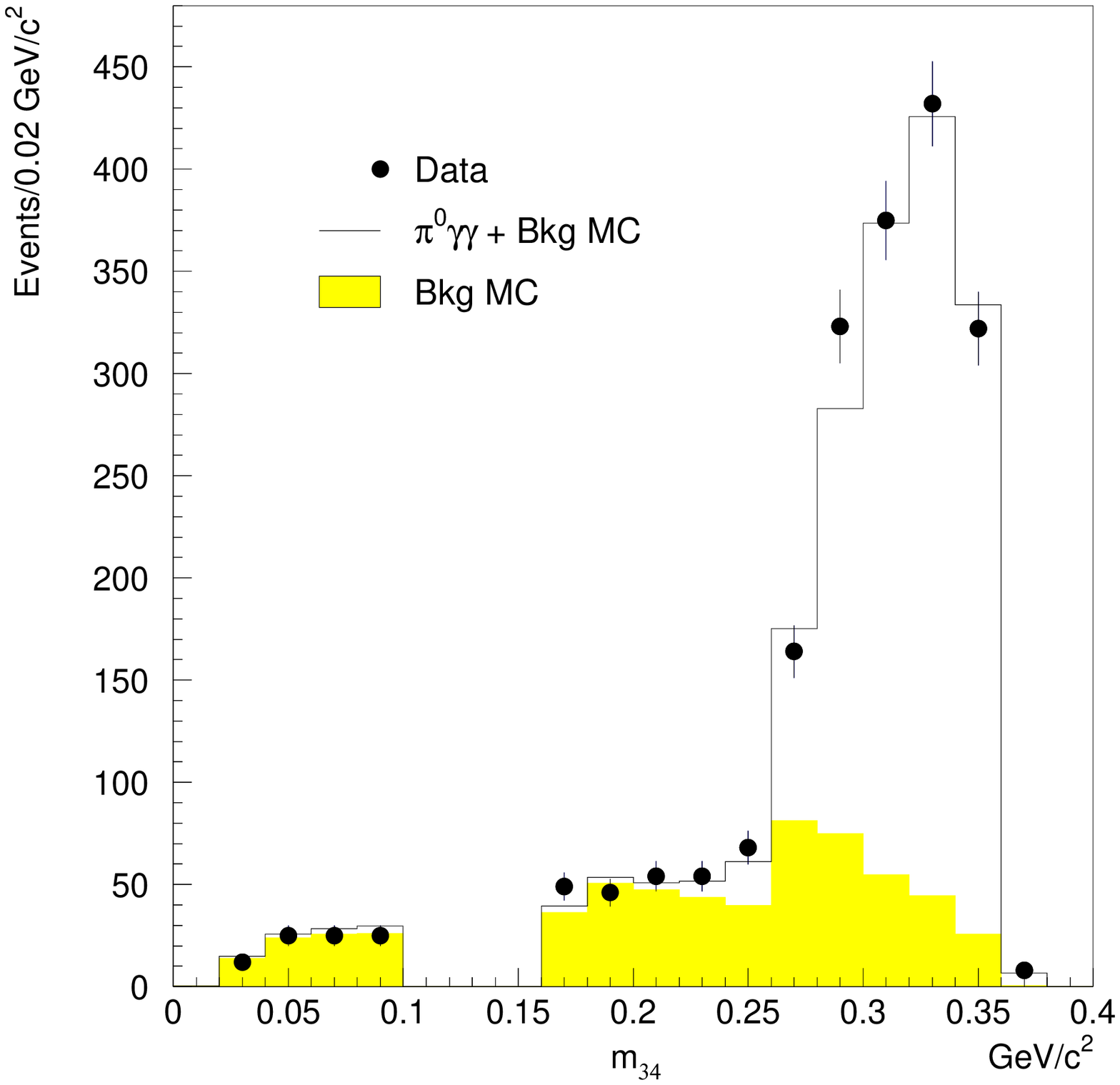}
	\caption{Distribution of the variable $Z_{Dalitz}$ used in the analysis of \ktopigg. 
           }
       \label{pic:ktopigg_zdalitz_plot}
 \end{figure}

After having obtained a sample of \ktopigg\ the resulting events were used to perform a likelihood fit using the model developed in \cite{DAmbrosio1997}
and the Dalitz variables $Z_{Dalitz} = M_{34}^{2}/\mkaon^{2}$  and $Y_{Dalitz} =\left(E_{\gamma3} - E_{\gamma4}\right)/\mkaon $ where $M_{34}$ is the invariant mass
of the two photons which are not daughters of the \pizero. The result is:
\begin{equation}
\av = -0.31 \pm 0.05_{stat} \pm 0.07_{syst} 
\end{equation}
The distributions of the Dalitz plot variables for both data and Monte Carlo, generated using the best fit value of \av, are shown in Figures \ref{pic:ktopigg_ydalitz_plot} 
and \ref{pic:ktopigg_zdalitz_plot}.

\subsection{\ktopieeg}
This decay, which can be though of as $\klong \to \pizero \gamma^{*} \gamma $, also sees contributions from the vector meson exchange process, and as such can also
be used to measure \av. Reconstruction begins by looking for two charged tracks and three neutral CsI clusters, which are not associated with any tracks.
Two of the three neutral CsI clusters are required to reconstruct with an invariant mass near the mass of the \pizero. The decay vertex obtained from the \pizero
reconstruction is then used to compute both $M_{ee\gamma}$ and $M_{ee\gamma\gamma\gamma}$. In this case the decay vertex obtained from the extrapolation of the
two charged tracks is not used as the two electron trajectories are nearly parallel and overlapping in this decay. Finally an additional constraint is applied that 
requires that no combination of
charged tracks and a photon may reconstruct near the invariant mass of the \pizero. After application of these and additional cuts \cite{Pi0eeg2007} we are left
with 139 events with an estimated background of approximately 14.4 events.

Isolating the decay \ktopizpizd, which has the same final state, and utilizing it as a normalization mode, we then obtain the branching ratio:
\begin{equation}
Br(\ktopieeg) = \left(1.62 \pm 0.17 \right) \times 10^{-8}
\end{equation}
which supersedes the older KTeV result \cite{Pi0eeg2001} which was affected by an older value of the \ktopizpizd\ branching ratio. This branching ratio measurement
also compares favorably to the value obtained using Chiral Perturbation Theory, which is $Br(\ktopieeg)_{ChPT} = 1.51 \times 10^{-8}$ \cite{Donoghue1997,Pi0eeg2007}
This value of the branching ratio also informs searches for \kltopiee, as the low value implies that this decay mode will not be a sizable contribution to the
background for \kltopiee. Furthermore, the distribution of $M_{\pizero ee}$ peaks well away from the kaon mass in this decay.

\begin{figure}[htb]     
	\includegraphics[scale=0.40]{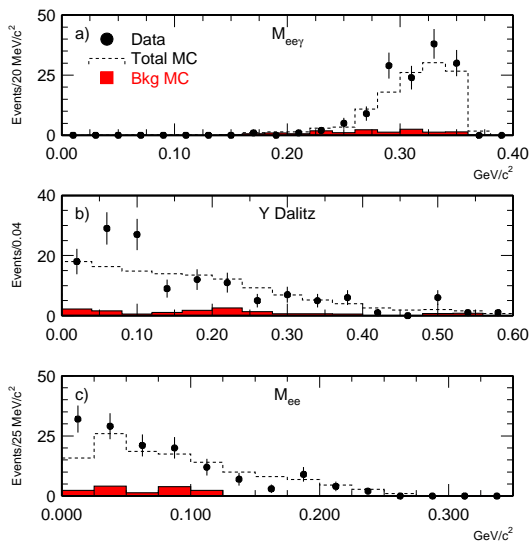}
	\caption{Distributions of the three Dalitz variables used in the analysis of \ktopieeg. 
           }
       \label{pic:ktopieeg_kin_plot}
 \end{figure}

The sample of \ktopieeg\ events can also be used to extract \av\ much in the same way as \ktopigg. Performing a likelihood fit using the model from Reference
\cite{DAmbrosio1997} and the kinematic variables $Z_{Dalitz} = M_{ee\gamma}^{2}/\mkaon^{2}$ , $Y_{Dalitz} =\left(E_{\gamma} - E_{ee}\right)/\mkaon $ and 
$Q_{Dalitz} = M_{ee}^{2}/\mkaon^{2}$. Note that $Z_{Dalitz}$ and $Y_{Dalitz}$ are analogues of the same variables in \ktopigg. The resulting fit produces 
\begin{equation}
\av = -0.76 \pm 0.16_{stat} \pm 0.07_{syst} 
\end{equation}
The distributions of the three kinematic variables, for data and Monte Carlo generated with the fitted value of \av, are shown in Figure
\ref{pic:ktopieeg_kin_plot}.

\section{The parity of the \pizero}
There has been considerable direct \cite{Samios1962} and indirect \cite{Panofsky1951,Chinowsky1954} evidence of the pseudo-scalar nature of the \pizero\ for some time. 
While the fact that the \pizero\ is a pseudo-scalar is not in doubt, it is interesting to
re-examine the issue by studying the decay \pitoeeee. In this decay, each electron-position pair act as an analyzer that reveals
the polarization of the virtual photon from the underlying decay $\pi^{0} \to \gamma^{*} \gamma^{*}$, and thus expose the parity of the
\pizero. 

Data from the 4-track trigger of KTeV's E799 phase were filtered in order to expose events consistent with the decays $\klong \to \pizero
\pizero \pizero_{DD}, \ \pizero_{DD} = \pitoeeee$. The first requirement was that two pairs of electromagnetic clusters in the CsI calorimeter,
unassociated with charged tracks, each reconstruct with an invariant mass consistent with a \pizero. The second requirement was that 
there are also four charged tracks in the detector whose energy deposition in the calorimeter are consistent with electrons/positrons. The momenta of the four tracks 
must yield a combined invariant mass also consistent with a \pizero. Finally, the invariant mass of the entire system was required to
be equal to that of the parent \klong. The summed momentum of all the daughter particles was also required to point back to the
kaon production target. After these cuts and additional cuts detailed in \cite{Pi0toeeee2008} there are 30511 candidate events with an estimated
background of 0.6\%. The main background was from  $ \klong \to \pizero_{D} \pizero_{D} \pizero $ where $\pizero_{D}= \pizero \to e^{+}e^{-}\gamma$

\begin{figure}[htb]     
	\includegraphics[scale=0.40]{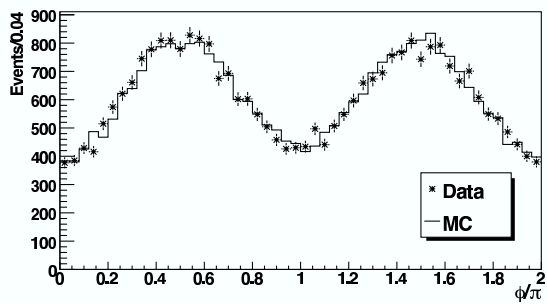}
	\caption{The distribution of angle between the planes defined by the
	two Dalitz pairs. There is a strong preference for the planes to be orthogonal, consistent with the parent being a
	pseudo-scalar particle. The non-zero number of events in the regions near $\phi=0,\pi,2\pi$ are a result of mispairing of the
        Dalitz pairs. 
           }
       \label{pic:phi_plot}
 \end{figure}

Once the \pitoeeee\ sample has been isolated, we can begin to analyze the kinematics of the decay. The plane of each Dalitz pair reveals
the polarization of the parent virtual photon. It is thus necessary to determine how to form the Dalitz pairs. 
We choose the pairing of the particles in a way such that $x_{1} < x_{2}$ and the product $x_{1} \times x_{2}$ is minimized, where $x_{1}$ and
$x_{2}$ are kinematic variables defined by:
\begin{equation}
x_{i} = \left(\frac{m_{e_{i}^{+}e_{i}^{-}}}{M_{\pizero}}\right)^{2}
\end{equation}
and are thus proportional to the invariant mass of each Dalitz pair. Note that there will be mispairings introduced using this method,
however this mispairing effect is minimized in the full treatment as detailed in \cite{Pi0toeeee2008} by utilizing a matrix element method.
Once the $e^{+}e^{-}$ pairs are formed, the angle $\phi$ is defined as the angle between the normal to the plane formed by
$e_{1}^{+}e_{1}^{-}$ and the plane formed by $e_{2}^{+}e_{2}^{-}$.  The resulting distribution of
the angle $\phi$ is shown in Figure ~\ref{pic:phi_plot}.

In addition to inspecting the distribution of the angle $\phi$, a detailed analysis of the form factor of the \pitoeeee\ decay was carried out. 
For this study, a maximum likelihood fit was carried out using the DIP model \cite{DAmbrosio1998} and various additional terms which allow for scalar and CPT
couplings \cite{Barker2003}. The fit yields a value for the DIP parameter 
\begin{equation}
\alpha = 1.3 \pm 1.0(stat) \pm 0.9(syst)
\end{equation}
which compares favorably to the standard
slope parameter from \pitoeeg.  In addition, the results of the fit constrain the presence of a scalar component in the decay to $ < 12.1\% \ \left(3.3\%\right)$  
at 90\% confidence when CPT is assumed to be violated (conserved).

In order to measure the branching ratio of this decay, we must also select an additional sample in order to obtain a normalization
sample. The decay $ \klong \to \pizero_{D} \pizero_{D} \pizero $ where $\pizero_{D}= \pizero \to e^{+}e^{-}\gamma$ is selected as a
normalization mode due to the similar final state. After all cuts,
141251 normalization events remain with a background of 0.5\%. The main background was $\klong \to \pizero \pizero \pizero_{DD}$ . Using the 
number of $\klong \to \pizero \pizero \pizero_{DD}$ events to the number of $ \klong \to \pizero_{D} \pizero_{D} \pizero $ events,
along with the previously measured value of the Dalitz branching ratio ( which is a leading contribution to the systematic error) we obtain:
\begin{equation}
Br(\pitoeeee) = \left(3.26 \pm 0.18 \right) \times 10^{-5}
\end{equation}
The above result also includes a detailed treatment of radiative corrections \cite{Barker2003}.
It should be noted that the KTeV collaboration intends to release a new measurement of $Br(\pitoeeg)$ in the near future, at which point the above branching ratio
can be recomputed and the systematic error reduced.

\section{Summary}
The KTeV collaboration has recently produced a suite of new limits on the lepton flavor violating decays \ktopimue, \ktopipimue\ and  \pitomue, all of which are either first
or current best limits. There were no expected backgrounds for all three decays, indicating that these limits, much like many other LFV searches, would be relatively straight
forward to improve in future experiments. In addition to LFV searches, the collaboration has also produced new measurements of the branching ratios and form factors for the
related decays \ktopigg\ and \ktopieeg, yielding two new estimates of the vector meson exchange amplitude \av. Finally, KTeV has produced a new, high statistics analysis of
the decay \pitoeeee\ and its decay characteristics.


\begin{thebibliography}{99}
\bibitem{EprimePRD2003} A. Alavi-Harati \emph{et al.} Phys. Rev. D \textbf{67} 012005 (2003)
\bibitem{Ahmad2002} Q.R. Ahmad \emph{et al.} Phys. Rev. Lett. \textbf{89}, 011301 (2002)
\bibitem{Langacker:1988vz}
  P.~Langacker, S.~Uma Sankar and K.~Schilcher,
  Phys.\ Rev.\  D {\bf 38}, 2841 (1988).
\bibitem{LFV2008} E. Abouzaid \emph{et al.}, Phys. Rev. Lett. \textbf{100}, 131803 (2008)  
\bibitem{Arisaka1998} K. Arisaka \emph{et al.}, Phys. Lett B \textbf{432}, 230 (1998)
\bibitem{Appel2000A} R. Appel \emph{et al.} Phys. Rev. Lett. \textbf{85}, 2450 (2000)
\bibitem{Appel2000B} R. Appel \emph{et al.} Phys. Rev. Lett. \textbf{85}, 2877 (2000)
\bibitem{AlaviHarati:2002eh}
  A.~Alavi-Harati {\it et al.} ,
  Phys.\ Rev.\ Lett.\  {\bf 90}, 141801 (2003)
\bibitem{Ambrose:1998us}
  D.~Ambrose {\it et al.}  ,
  Phys.\ Rev.\ Lett.\  {\bf 81}, 5734 (1998)
\bibitem{Spadaro2008} T. Spadaro, talk at Heavy Quark and Leptons 2008, http://hql08.ph.unimelb.edu.au/
\bibitem{Cirigliano:2007ga}
  V.~Cirigliano and I.~Rosell,
  JHEP {\bf 0710}, 005 (2007)
\bibitem{Testa:2008xz}
  M.~Testa  ,
  arXiv:0805.1969 [hep-ex].
\bibitem{Masiero:2005wr}
  A.~Masiero, P.~Paradisi and R.~Petronzio,
  Phys.\ Rev.\  D {\bf 74}, 011701 (2006)
\bibitem{Ahmed:2001eh}
  M.~Ahmed {\it et al.}  ,
  Phys.\ Rev.\  D {\bf 65}, 112002 (2002)
\bibitem{Hisamatsu:2007zza}
  Y.~Hisamatsu,
  Eur.\ Phys.\ J.\  C {\bf 52}, 477 (2007).
\bibitem{Mu2e2007} R.M. Carey, \emph{et al}, Fermilab TM-2396-AD-E-TD (2007)
\bibitem{Dohmen:1993mp}
  C.~Dohmen {\it et al.}  ,
  Phys.\ Lett.\  B {\bf 317}, 631 (1993).
\bibitem{Batley:2003mu}
  J.~R.~Batley {\it et al.}  ,
  Phys.\ Lett.\  B {\bf 576}, 43 (2003)
\bibitem{Gabbiani2002} F. Gabbiani and G. Valencia,  Phys. Rev. D \textbf{66}, 074006 (2002)
\bibitem{Pi0gg2008} E. Abouzaid \emph{et al.}, Phys. Rev. D \textbf{77}, 112004 (2008)
\bibitem{Pi0gg1999} A. Alavi-Harati \emph{et al.}, Phys. Rev. Lett. \textbf{83}, 917 (1999)
\bibitem{DAmbrosio1997} G. D'Ambrosio and J. Portoles,   Nuclear Phys. B \textbf{492}, 417 (1997)
\bibitem{Pi0eeg2007} E. Abouzaid \emph{et al.}, Phys. Rev. D \textbf{76}, 052001(2007)
\bibitem{Pi0eeg2001} A.Alavi-Harati \emph{et al.}, Phys. Rev. Lett. \textbf{87}, 021801 (2001)
\bibitem{Donoghue1997} J.F. Donoghue and F. Gabbiani,  Phys. Rev. D \textbf{56}, 1605 (1997)
\bibitem{Lai2002} A. Lai \emph{et al.} Phys. Lett. B \textbf{536}, 229 (2002)
\bibitem{Samios1962} N. P. Samios \emph{et al.}, Phys. Rev. \textbf{126} 1844 (1962) 
\bibitem{Panofsky1951} W. Panofsky \emph{et al.} Phys. Rev. \textbf{81} 565 (1951)
\bibitem{Chinowsky1954} W. Chinowsky and J. Steinberger Phys. Rev. \textbf{95} 1561 (1954)
\bibitem{Pi0toeeee2008} E. Abouzaid \emph{et al.}, Phys. Rev. Lett. \textbf{100}, 182001 (2008)
\bibitem{DAmbrosio1998} G. D'Ambrosio \emph{et al.}, Phys. Lett. B \textbf{423} 385 (1998)
\bibitem{Barker2003} A.R Barker \emph{et al.} Phys. Rev. D \textbf{67}, 033008 (2003)

\end{thebibliography}
\end{document}